\documentclass[aps,prl,twocolumn,groupedaddress,showpacs]{revtex4}
\usepackage{amsmath,amssymb,graphicx}

\begin{document}

\title{
Observation of the Fano-Kondo Anti-Resonance
in a Quantum Wire \\
with a Side-Coupled Quantum Dot
}

\author{Masahiro~Sato,
Hisashi~Aikawa, Kensuke~Kobayashi,
Shingo~Katsumoto, and Yasuhiro~Iye}
\affiliation{Institute for Solid State Physics, University of Tokyo,
5-1-5 Kashiwanoha, Chiba 277-8581, Japan}
\date{\today}

\begin{abstract}
We have observed the Fano-Kondo anti-resonance in
a quantum wire with a side-coupled quantum dot.
In a weak coupling regime,
dips due to the Fano effect appeared.
As the coupling strength increased, conductance in the regions
between the dips decreased alternately.
From the temperature dependence and the response to the magnetic field,
we conclude that the conductance reduction is due to the Fano-Kondo anti-resonance.
At a Kondo valley with the Fano parameter $q\approx 0$, the phase shift is locked to $\pi/2$ against the gate voltage when the system is close to the unitary limit in agreement with theoretical predictions by Gerland {\it et al.}~[Phys. Rev. Lett. {\bf 84}, 3710 (2000)].
\end{abstract}
\pacs{72.15.Qm, 73.21.La, 73.23.Hk}

\maketitle 
The experimental realization of the Kondo effect~\cite{Kondo} in a semiconductor quantum dot (QD) has
opened up a new human-made stage to investigate many body effects~\cite{gordon}.
In a previous paper, we have shown that
spin-scattering at a QD, which means the creation of a spin-entangled state
between a localized spin and a conduction electron,
leads to quantum decoherence~\cite{Aikawa, Koenig}.
At the same time, however, such an entangled state is a
starting point to build a Kondo singlet~\cite{Yosida}, in which
the coherence of electrons is expected to be recovered.

\begin{figure}
\includegraphics{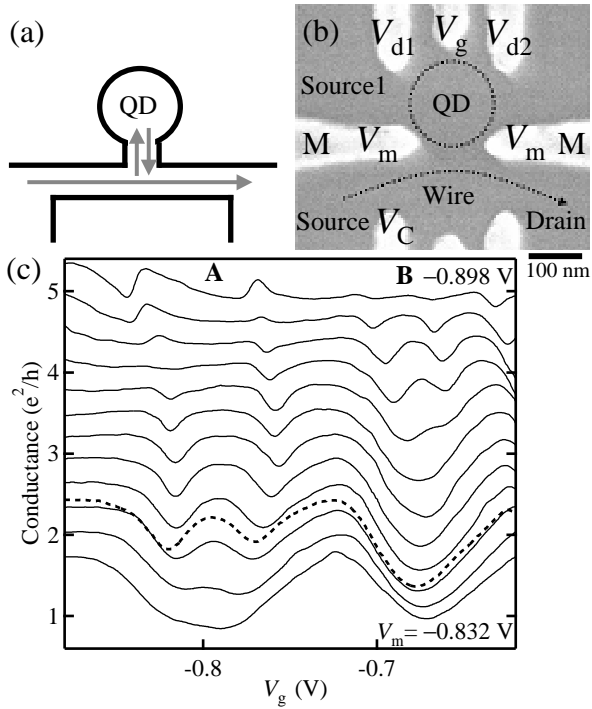}
\caption{
(a) Schematic diagram of a stub-resonator.
(b) Scanning electron micrograph of the device. The white areas are metallic gates
made of Au/Ti. 
The dot and the wire are indicated by dotted lines.
Conduction between Source1 and Drain is used to measure the direct transport through
the dot, while that between Source and Drain gives the main results of this paper.
(c) Conductance of the wire measured at several coupling strength
tuned by $V_{\rm m}$. The base temperature is 30~mK.
The step in $V_{\rm m}$ is 6~mV and the data are offset by 0.3$e^2/h$ for clarity.
The broken line shows the data at $V_{\rm m}=-$0.846~V shown in Fig.~\ref{Fig2}.
}
\label{Fig1}
\end{figure}

The direct evidence for the coherence of the Kondo state is the interference
effect through a QD in the Kondo state.
Ji {\it et al}. measured the phase shift of electrons through a dot in the Kondo state
in an Aharonov-Bohm (AB) ring~\cite{Ji}, and
found that the phase shift significantly varies even
on the Kondo plateau of conductance, which may suggest the breakdown of
the Anderson model.
However, the difficulty in measuring the phase shift in AB geometry is
pointed out~\cite{Aharony}, and experiments in simpler
structures such as a stub-resonator (a schematic diagram is shown in Fig.~\ref{Fig1}(a))~\cite{Kang,Thimm} are desired.

Some of the present authors have reported the observation of the Fano anti-resonance
in a quantum wire with a side-coupled QD~\cite{Kobayashi_antifano},
which is a kind of a stub-resonator.
The Fano effect~\cite{Fano} is a consequence of interference
between a localized state and a continuum, which correspond to
a state in the QD and that in the wire, respectively.
It appears as a characteristic line shape in the conductance $G$ as
\begin{equation}
G (\epsilon) \propto (\epsilon + q)^2/(\epsilon^2 + 1),
\label{eqFano}
\end{equation}
where $\epsilon$ is the energy difference from the resonance position
normalized with the width of the resonance,
and $q$ is Fano's asymmetric parameter~\cite{Kobayashi}. 
The Fano parameter represents the degree of distortion and $q=0$ corresponds to
an anti-resonance dip.
The Fano-Kondo effect -- the Fano effect which appears in the Kondo cloud --
is hence sensitive to coherence and phase shifts~\cite{Hofstetter}.
In this Letter, we report the observation of the Fano-Kondo anti-resonance
in a quantum wire
with a side-coupled QD.
Phase shift locking to $\pi/2$ is deduced from the analysis of
the line shape of the anti-resonance.

Our device consists of a quantum dot and a quantum wire
difined in a two-dimensional electron gas 
(2DEG, sheet carrier density $3.8 \times 10^{15}~{\rm m^{-2}}$,
mobility $80~{\rm m^2/Vs}$)
formed at a GaAs/AlGaAs hetero-structure.
Au/Ti metallic gates were deposited as shown in Fig.~\ref{Fig1}(b).
A QD is defined by the upper three gates and the two gates marked as ``M" (M-gates).
The lower three gates adjust the conductance of the wire.
The middle of the upper gates is used to control 
the potential of the dot (gate voltage $V_{\rm g}$)
and M-gates tune the coupling between the dot and the wire (gate voltage $V_{\rm m}$).
It was confirmed that the system works as a quantum wire with no quantum dot
when M-gates are pinched-off.
Tuning the coupling strength by M-gates induces the
electrostatic potential shift of the dot,
whereas the potential shift by $V_{\rm g}$ causes little change
in the coupling strength.
Accordingly, the potential shift by $V_{\rm m}$ can be compensated
by $V_{\rm g}$ (approximately $\varDelta V_{\rm g}=-4.5\varDelta V_{\rm m}$
in the present sample).
For the sake of a simpler description, 
we henceforth redefine $V_{\rm g}\equiv V_{\rm g}({\rm raw})-4.5(V_{\rm m}+0.846)$.
In order to attain a large value of Kondo temperature $T_{\rm K}$,
the device was designed to 
make the dot size smaller than that used in Ref.~\cite{Kobayashi_antifano}.
The dot is placed close to the wire
to avoid temperature dependence associated with a change of the coherence 
length, which is unfavorable for the present study.
Although the proximity of the dot to the wire would cause a Fano-charging mixing effect~\cite{Johnson},
it is less severe in a comparatively strong coupling regime explored here.
The sample
was cooled in a dilution refrigerator down to
30~mK and was measured by standard lock-in techniques in a
two-terminal setup.

In order to obtain the relevant parameters of the dot,
we first measured the direct charge transport through the dot (from ``Source1" to
``Drain" in Fig.~\ref{Fig1}(b))
by slightly opening $V_{\rm d1}$ and by closing $V_{\rm d2}$ and $V_{\rm c}$.
The average level spacing $\varDelta$ obtained from excitation spectroscopy is about 0.3~meV,
which gives the dot diameter as about 170~nm and 
the total number of electrons
as about 90.
The electron temperature estimated from the widths of resonance peaks
followed the fridge temperature down to 100~mK and severe saturation occurred below that.
Then the direct connection was pinched off by $V_{\rm d1}$
and the wire was opened by $V_{\rm c}$.
The wire conductance far from (anti-)resonances was adjusted to be around
$G_q\equiv 2e^2/h$, {\it i.e.}, at the first step of the conductance staircase.

\begin{figure}
\includegraphics{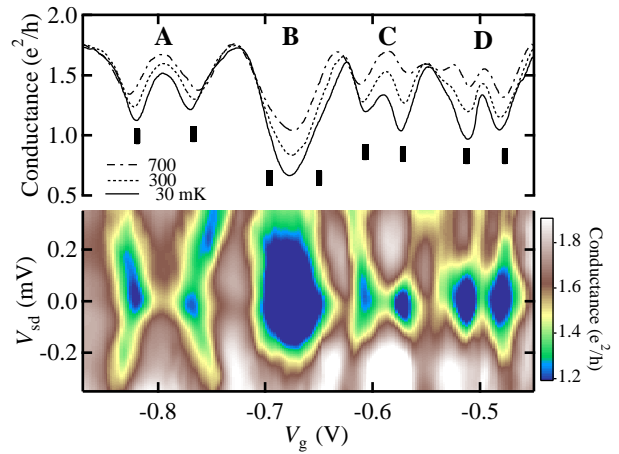}
\caption{
Upper: Zero-bias conductance of the quantum wire at three different temperatures
as a function of the gate voltage $V_{\rm g}$ at $V_{\rm m}=-0.846$~V.
The bars indicate the positions of Coulomb ``dips"
and A-D indicate the Coulomb valleys where the Kondo effect emerges.
Lower: Color scale plot of the conductance 
on the $V_{\rm g}$-$V_{{\rm sd}}$ plane at 30~mK.
Both the upper and the lower panels use the same $V_{\rm g}$ axis.
}
\label{Fig2}
\end{figure}

Figure \ref{Fig1}(c) shows the wire conductance against $V_{\rm g}$ 
at various coupling strength adjusted by $V_{\rm m}$.
In a weak coupling regime (at the top of Fig.~\ref{Fig1}(c)), 
Coulomb ``dips" appear, reproducing the
previous result~\cite{Kobayashi_antifano}.
This is due to the destructive interference, {\it i.e.}, the Fano anti-resonance.
These dips are regularly placed and the averaged period is the same as 
that of the Coulomb oscillation in the direct transport.
This indicates that the dips are originated from the QD.
As the coupling strength increases,
conductance between the dips (Coulomb valleys)
decreases alternately.
These reductions connect two neighboring
dips into one (in regions A and B).
We have observed four such valleys (as will be shown in Fig.~\ref{Fig2}) where the conductance decreases
as the coupling strength increases.
We next confirm that these reductions are due to the Fano-Kondo
anti-resonance, then discuss
the obtained information.

The upper panel of Fig.~\ref{Fig2} shows the zero-bias wire conductance as a function of 
$V_{\rm g}$ at three different temperatures
for a medium coupling strength.
The closed bars indicate the positions of the Coulomb dips.
For valley B, the positions cannot be resolved directly
and are extrapolated from the positions at $V_{\rm m}=-0.886$~V.
In the four regions indicated as A-D, conductance decreases with
decreasing temperature.
The Kondo effect for spin 1/2 emerges for a QD with odd number of electrons.
Hence it usually appears alternately at Coulomb valleys
as we observed at Kondo valleys A-D.

The distances between the Coulomb dips are shorter at the Kondo valleys than those 
at the neighboring valleys.
The level spacing $\varDelta$ obtained from this difference of the distances
is in good agreement with $\varDelta$
obtained from excitation spectroscopy.
This means the simple ``spin-pair" picture is holded and
the conditions
for the Kondo effect are fulfilled at valleys A-D.

The lower panel of Fig.~\ref{Fig2} is a color scale plot of the conductance on
the plane of $V_{\rm g}$ and the source-drain bias voltage $V_{{\rm sd}}$.
The conductance drops around $V_{{\rm sd}}=0$~mV at the Kondo valleys.
Note that $V_{{\rm sd}}$ is not directly applied
to the dot but to the wire.
The wire by itself can show non-linear conductance~\cite{Kristensen}.
In this experiment, however, the non-linearity of the wire is small, because the wire is under 
the plateau condition and the scale of $V_{{\rm sd}}$ is magnitude smaller than that in the previous reports~\cite{Kristensen}.
Furthermore, the response to $V_{\rm g}$ and the reversed Coulomb diamond like structure
(not well resolved in Fig.~\ref{Fig2}) indicate
that the zero-bias ``dips" originate from the Kondo effect of the dot.
The charging energy $U$ estimated from the reversed Coulomb diamond
ranges from 0.3 to 0.5~meV.
For a quantitative comparison with existing theories, we should be careful
to multi-level effect, which is usually ignored by taking the limit of
infinite $\varDelta$ (or charging energy). Such ``single-level" approximation
is expected to hold when the separation of Coulomb peaks (dips) is
clear. 
We therefore concentrate on dip A, which shows the clearest seperation.

\begin{figure}
\includegraphics[width=\linewidth]{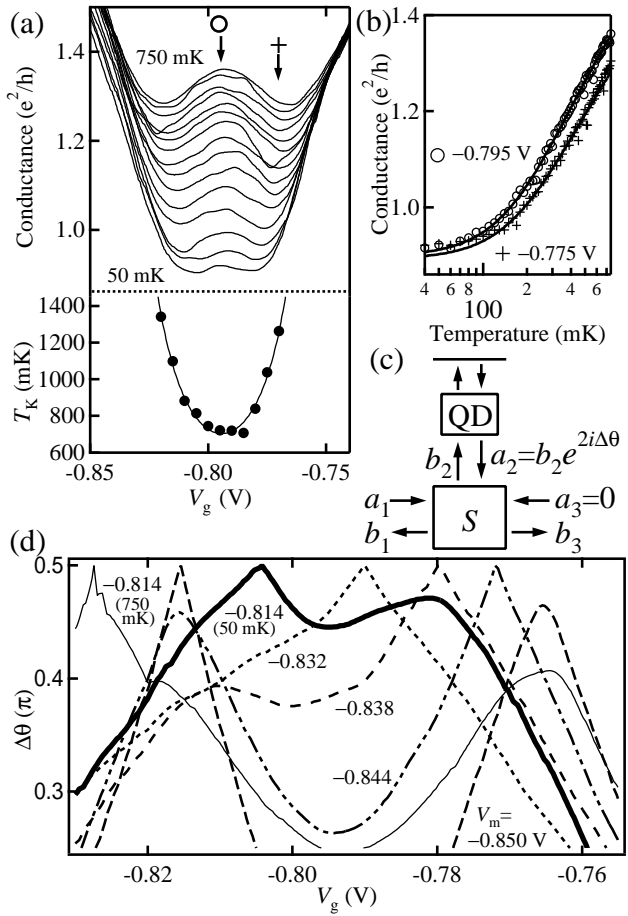}
\caption{
(a) Upper: Conductance 
at temperatures from 750~mK to 50~mK with the step of 50~mK at $V_{\rm m}=-0.814$~V.
Lower: Kondo temperatures $T_{\rm K}$ 
obtained from the temperature dependence.
(b) Examples of the fitting to obtain $T_{\rm K}$.
The gate voltages adopted here are indicated by arrows in (a).
(c) Schematic diagram of a simple quantum circuit model of the system. The factor 2 of $\varDelta\theta$
in $a_2$ is due to the reflection.
(d) Phase shift of the QD obtained for the data in (a) at $V_{\rm m}=-0.814$~V and 50~mK (thick line), 750~mK (thin line), and for the data in Fig.~\ref{Fig1}(c) at $V_{\rm m}=-$0.832~V, $-$0.838~V, $-$0.844~V, $-$0.850~V (dotted lines), based on the model shown in (c). 
The data folding at $\pi/2$ is due to taking plus of the double sign in \eqref{complex}.
}
\label{Fig3}
\end{figure}

The upper panel of Fig.~\ref{Fig3}(a) shows detailed temperature dependence of 
$G(V_{\rm g})$ at Kondo valley A.
We obtained the Kondo temperature $T_{\rm K}$ from
the form~\cite{costi}:
\begin{equation}
G(T)=G_0-G_1\left({T_{\rm K}'^2}/({T^2+T_{\rm K}'^2})\right)^s,
\label{tdep}
\end{equation}
where $G_{1}$, $T_{\rm K}'\equiv T_{\rm K}/\sqrt{2^{1/s}-1}$, and $s$ are fitting parameters.
$G_0$ was fixed to 1.8$e^2/h$, the wire conductance far from the anti-resonance.
Examples of the fitting are shown in Fig.~\ref{Fig3}(b). 
The data below 120~mK are not taken
into the fitting considering the electron temperature saturation.
From the fitting, $s=0.25\pm0.04$ was obtained,
which is in accordance with the prediction for spin 1/2 impurities.
The obtained $T_{\rm K}$ are plotted against $V_{\rm g}$
in the lower panel of Fig.~\ref{Fig3}(a).
$T_{\rm K}$ depends parabolically on $V_{\rm g}$ with the bottom around the mid-point of 
the Kondo valley just like the previous reports~\cite{vanderwiel}.
This dependence agrees well with 
$T_{\rm K}=\sqrt{\varGamma U}\exp(\pi\epsilon(\epsilon+U)/\varGamma U)/2$,
where $\varGamma$ is the dot-wire coupling and $\epsilon$ is the single
electron level measured from the Fermi level.
We obtained $U=0.39\pm0.03$~meV and $\varGamma=0.30\pm0.02$~meV
by fitting the above function to $T_{\rm K}$.

\begin{figure}
\includegraphics{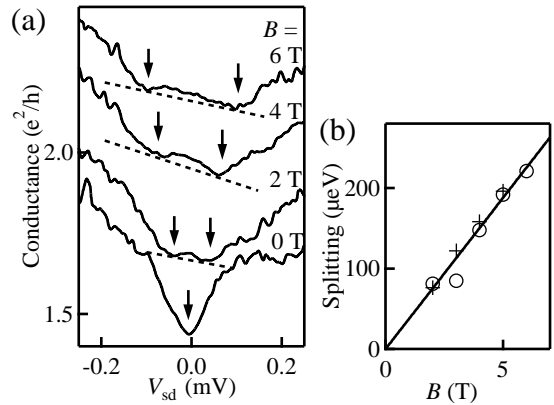}
\caption{
(a) Conductance versus $V_{{\rm sd}}$ at Kondo valley A
under various magnetic fields up to 6~T. 
The arrows indicate the positions of
the conductance dips, which are determined by fitting two Gaussian curves.
The broken lines are eye guides to show the baselines.
(b) Distance between two dip positions versus the magnetic field. 
The data were taken at Kondo valleys A (open circles) and C (crosses) in Fig.~\ref{Fig2}.
The line corresponds to $|g|=0.33$.
}
\label{Fig4}
\end{figure}

Figure \ref{Fig4}(a) shows the splitting of the zero-bias conductance dip under the external magnetic
field parallel to the 2DEG.
The splitting is proportional to
the field as shown in Fig.~\ref{Fig4}(b)
and attributed to the Zeeman splitting of the Kondo anomaly.
From the slope of the fitted line, we obtain the g-factor of electrons as $|g|=0.33$,
which is close to that reported for GaAs/AlGaAs 2DEG~\cite{Jiang}.

So far we have confirmed that the conductance reduction
in regions A-D in Fig.~\ref{Fig2} is due to the Fano-Kondo anti-resonance.
In the side-coupled geometry, only destructive interference can cause
the conductance reduction.
The observation of the Kondo effect through the pure interference
effect manifests that electron transport through the Kondo cloud
is coherent as predicted~\cite{Kang}.

In Fig.~\ref{Fig1}(c),
two dips merges into one in regions A and B.
This means the system almost reaches the unitary limit~\cite{vanderwiel}.
This is consistent with the obtained $T_{\rm K}$ shown in the lower panel of Fig.~\ref{Fig3}.
The residual conductance at the ``unitary limit" Kondo valleys
({\it e.g.}, 0.7$e^2/h$ at valley B in Fig.~\ref{Fig2})
is probably due to
the finite area between the wire
and the dot, which
works as a resonator and spoils the perfect
reflection~\cite{Thimm}, while the phase shift is unchanged
because $q \approx 0$.

When the coupling is weak (at the top
of Fig.~\ref{Fig1}(c)), the two dips at both sides of 
valley A have asymmetric Fano line shapes ($q \neq 0$).
It has been anticipated that
symmetry of transport through a QD is dominated
by a few states with an anomalously strong coupling
to an electrode (strong coupling states, SCSs)~\cite{Silvestrov}.
In a weak coupling regime, the $q$'s of the two dips at both sides of valley A are hence determined by an SCS.
As the coupling strength increases, however, the values of $q$ 
approach to zero
indicating that the coupling strength is
renormalized and the Kondo state
takes over the SCS.

We now discuss the phase shift of Kondo state, which can be obtained from the analysis
along the scattering form of conductance.
A simple model of the system : a three branch S-matrix and a reflector (the QD), 
is shown in Fig.~\ref{Fig3}(d).
Here we take the S-matrix simply as
\begin{equation}
\begin{pmatrix}
b_1\\b_2\\b_3
\end{pmatrix}
=
\begin{pmatrix}
0&1/\sqrt{2}&1/\sqrt{2}\\
1/\sqrt{2}&-1/2&1/2\\
1/\sqrt{2}&1/2&-1/2
\end{pmatrix}
\begin{pmatrix}
a_1\\a_2\\a_3
\end{pmatrix}.
\label{smatrix}
\end{equation}
Assuming the absense of reflection from the electrodes we put $a_3=0$.
And taking the QD as a transmitter plus a perfect reflector,
we put $a_2=b_2\exp(2i\varDelta\theta)$ and $a_1=1$ for unitary input.
Then the phase shift $\varDelta\theta$ is easily obtained as
\begin{equation}
\varDelta\theta={\rm arg}\left(\frac{\sqrt{2}-2b_3}{b_3-\sqrt{2}}\right),
\;\;
{\rm arg}(b_3)=\pm\cos^{-1}\left(\frac{3|b_3|}{2\sqrt{2}}\right).
\label{complex}
\end{equation}
Here $b_3$ is the complex transmittance and related with the conductance through the Landauer
formula as $G=G_0'\text{(constant)}+(e^2/h)|b_3|^2$.
The specific form of \eqref{smatrix} does not spoil the generality.
The time-reversal symmetry, the unitarity and $q=0$ impose strong constraints on the S-matrix
and the residual freedom has very little effect.

Figure \ref{Fig3}(d) shows thus calculated $\varDelta\theta(V_{\rm g})$.
Here we take plus of the double sign in \eqref{complex}, which causes
the virtual folding at $\pi/2$.
This is due to the special condition of $q=0$ and the folding does not disturb the
confirmation of locking to $\pi/2$.
As temperature decreases from 750~mK to 50~mK and the coupling becomes stronger 
from $V_{\rm m}=-0.850$~V to $-0.814$~V, the phase shift variation approaches to 
the stationary line of $\pi/2$ at the Kondo valley.
This is in accordance with the existing theories~\cite{gerland}
and in contrast to the previous result~\cite{Ji}, where
only locking to $\pi$ at pre-Kondo region was observed.
The difference may come from the width of the energy levels~\cite{gerland}, or
from the geometry of the interferometer, 
though we have no conclusive opinion at present.

In Fig.~\ref{Fig2}, zero-bias anomalies appear
while no voltage was directly applied to the dot.
This indicates that the Kondo cloud spreads into the wire and is affected by $V_{\rm sd}$.
This is reasonable considering the size of the Kondo cloud
$\hbar v_{\rm F}/k_{\rm B}T_{\rm K}$ ($v_{\rm F}$: Fermi velocity), which exceeds 2~$\mu$m
for $T_{\rm K}$ less than 1~K~\cite{Simon}.
The present structure provides means
to investigate the size of the Kondo cloud.

In summary, we have observed the Fano-Kondo anti-resonance
in a quantum wire with a side-coupled dot.
Phase shift locking to $\pi/2$ is deduced from 
the line shape of the anti-resonance.

This work is supported
by a Grant-in-Aid for Scientific Research and by a Grant-in-Aid for
COE Research from the Ministry
of Education, Culture, Sports, Science, and Technology of
Japan.

\end{document}